\documentclass[aps,pra,reprint]{revtex4-1}

\usepackage{amsmath}
\usepackage{graphicx} 
\usepackage[sort&compress]{natbib}
\usepackage{colortbl}
\definecolor{lightgray}{gray}{0.9}

\begin{document}


\title{Free energy calculation of mechanically unstable but \\ dynamically stabilized bcc titanium}


\author{Sara Kadkhodaei}
\email[]{To whom correspondence should be addressed; Email: sara\_kadkhodaei@brown.edu}
\author{Qi-Jun Hong}
\author{Axel van de Walle}
\affiliation{Box D, School of Engineering, Brown University, 184 Hope street, Providence, Rhode Island 02912, USA}
\date{\today}
\begin{abstract}
The phase diagram of numerous materials of technological importance features high-symmetry high-temperature phases that exhibit phonon instabilities. Leading examples include shape-memory alloys, as well as ferroelectric, refractory, and structural materials. The thermodynamics of these phases have proven challenging to handle by atomistic computational thermodynamic techniques, due to the occurrence of constant anharmonicity-driven hopping between local low-symmetry distortions, while maintaining a high-symmetry time-averaged structure.\\
To compute the free energy in such phases, we propose to explore the system's potential-energy surface by discrete sampling of local minima by means of a lattice gas Monte Carlo approach and by continuous sampling by means of a lattice dynamics approach in the vicinity of each local minimum. Given the proximity of the local minima, it is necessary to carefully partition phase space by using a Voronoi tessellation to constrain the domain of integration of the partition function, in order to avoid double counting artifacts and enable an accurate harmonic treatment near each local minima. We consider the bcc phase of titanium as a prototypical examples to illustrate our approach.
\end{abstract}


\maketitle

\section{Introduction}
\begin{figure}[h!]
\begin{center}
\includegraphics[width=\columnwidth]{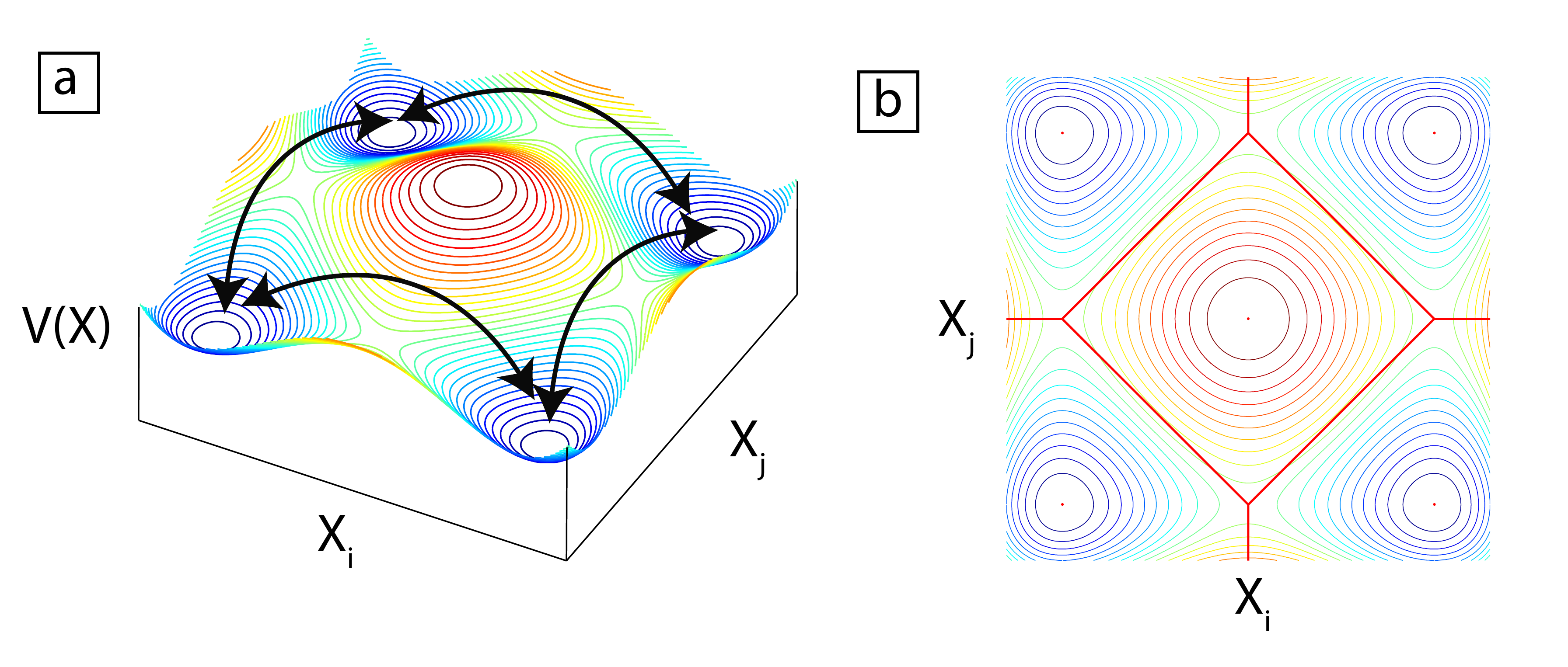}
\caption{Partitioning the configuration space into different regions. (a) Schematic of potential energy including multiple local minima around the high symmetry point. (b) Potential-energy contour represented on the configuration space which is partitioned into Voronoi tessellation. Each Voronoi cell is associated with a configuration.}
\label{fig:multiple_local_min}
\end{center}
\end{figure}

{\label{sec:intro}}
Phase diagrams are widely used as a powerful tool to predict the equilibrium state of physical systems. In recent years, computational methods have extensively contributed to the determination of thermodynamic data~\cite{comp-3}. The underlying assumption in many of the commonly used frameworks for phase diagram computation, such as cluster expansion~\cite{Sanchez84,vandewalle2002,fultz:progrev,clus-3} and computational thermodynamics~\cite{calph-1,calph-2,calph-3,calph-4}, is that all structural phases, even if they are observed at elevated temperatures, are mechanically stable. This ``lattice stability" assumption, however, reduces the applicability of the aforementioned frameworks, because the crystal structure of many high-temperature phases exhibit mechanical instabilities. These instabilities manifest themselves by phase transitions to lower-symmetry structures at low temperatures and can be readily identified by lattice dynamics calculations based on a harmonic model centered about the high-symmetry high-temperature structure~\cite{instab-dyn-1,instab-dyn-2,instab-dyn-3,instab-dyn-4,instab-dyn-5,instab-dyn-6,instab-dyn-7,instab-dyn-8,instab-hydride-1,instab-hydride-2,instab-oxide}. 
For example, many transition metals~\cite{instab-dyn-1,instab-dyn-2,instab-dyn-3,instab-dyn-4,instab-dyn-5}, their alloys~\cite{instab-dyn-6,instab-dyn-7,instab-dyn-8}, accompanied by their hydrides~\cite{instab-hydride-1,instab-hydride-2} and oxides~\cite{instab-oxide}, all reveal high-symmetry phases at elevated temperatures with mechanical instabilities. These mechanical instabilities are also common among shape memory alloys~\cite{instab-shapeMemory-1,instab-shapeMemory-2}, refractory oxides,~\cite{instab-refractory-1,instab-refractory-2} 
and ferroelectric materials~\cite{PbTe:ferroelectric,BiS2:ferroelectric}. 
In some cases (such as fcc W), the mechanical instability is such that the phase simply does not exist in nature~\cite{unstab-1,free-calc-unst-1,free-calc-unst-2,instab-dyn-8}. But in many cases (such as bcc Ti) the phase is ``dynamically stabilized " at high temperatures, thanks to entropy contributions arising from constant hopping between local low-symmetry distortions of a high- symmetry structure~\cite{Zener,instab-dyn-1}.
\\
Standard lattice dynamic calculations, such as (quasi)-harmonic models, are inadequate for the free energy calculation of such phases, because the energy surface becomes nonconvex along unstable modes and introduces nonphysical divergence in the calculation of the free energy. To calculate the free energy of such phases accurately, a practical and efficient framework has to be devised to account for anharmonic vibrational effects, which play a crucial role in the stabilization of high-temperature phases with low-temperature mechanical instabilities. 
\\
A number of solutions have been proposed so far~\cite{free-calc-dyn-1,MD,Debye,self-cons-phonon-1,self-cons-phonon-2,self-cons-phonon-3,self-cons-phonon-4,eff-Ham-1,eff-Ham-2,eff-Ham-3,eff-Ham-4,marianetti:slave,PbTe:slave,compressive_sensing,TU_TILD}, although some of them are not free of certain drawbacks. Brute force \textit{ab initio} molecular dynamics (AIMD) calculations can deliver free energy changes as a function of temperature, but obtaining absolute free energy (as would be needed to predict phase transitions) remains a challenge,
due to the fact that a simple reference state with a known free energy is unavailable.
Thermodynamic integration (TI) approaches use AIMD in order to calculate the anharmonic free energy by employing quasiharmonic calculations as a reference~\cite{TU_TILD}. 
However, when anharmonic effects are strong, computational requirements can become intractable for TI approaches (see Sec. \ref{sec:discussion}). 
Self-consistent phonon theories provide more computationally inexpensive avenues and have been successful in predicting effective phonon frequencies and free energies in a range of systems~\cite{self-cons-phonon-2,self-cons-phonon-3,self-cons-phonon-4}. However, these methods fundamentally rely
on the assumption of the existence of an effective harmonic model. If this assumption is inappropriate, there is no systematic avenue to improve the accuracy of the model.
Effective-Hamiltonians approaches~\cite{eff-Ham-1,eff-Ham-2,eff-Ham-3,eff-Ham-4,marianetti:slave,PbTe:slave}, which explicitly parametrize the system's anharmonic energy surface, do offer systematically improvable models without significant \textit{a priori} assumptions. Although these approaches have proven to be powerful tools to investigate phase-transition phenomena, the task of parametrizing the relevant anharmonic degree of freedom can become daunting as the range of interactions considered is increased. Systematic and rigorous approaches have been proposed recently, which significantly improve the efficiency of the task of incorporating anharmonic effects by selecting the physically important degrees of freedom in the lattice dynamics model, exploiting the compressive sensing techniques~\cite{compressive_sensing}.
\\
In this paper, we propose a systematically improvable method that simplifies the anharmonic energy surface parametrization process by breaking the problem into a collection of simpler tasks. Our approach is based on coarse graining of the partition function, subjected to an innovative Voronoi partitioning of the configuration space, which enables the use of a piecewise quadratic approximation to the system's potential-energy surface in conjunction with a cluster expansion approach.
We suggest the name ``piecewise polynomial potential partitioning'' or ``$\text{P}^\text{4}$'' for the proposed method.
As shown in Fig. \ref{fig:multiple_local_min}, a local exploration of multiple minima around the high-symmetry structure on the potential-energy surface is accomplished by partitioning the configuration space into multiple corresponding regions. 
The partitioning scheme ensures that the harmonic approximation is not employed beyond its range of validity and eliminates unphysical divergences in the calculated harmonic free energy.
Each local harmonic model can be constructed by means of well-established lattice dynamics techniques and we devise a Monte Carlo scheme to calculate the associated free energy contribution that accounts for the region boundaries. The free energy contributions arising from hopping between the different regions is accounted for by means of the cluster expansion formalism in conjunction with lattice gas Monte Carlo simulations.
The advantage of proceeding in this way is that the very complex anharmonic system is reduced to two, nested, simple classes of models: A lattice gas and an array of constrained harmonic models, for which automated construction methods are available~\cite{atat1,atat2,atat3}.
\\
The application of the $\text{P}^\text{4}$ method is to describe the equilibrium state of a phase with its thermodynamic properties, when opposed to kinetic models, as needed to build the phase diagram of the system. This application is most advantageous when the lattice vibrations become anharmonic or even``harmonically unstable".
In Sec. \ref{sec:p4} a general description of the $\text{P}^\text{4}$ method is presented employing bcc Ti as an illustrative example. In Sec. \ref{sec:result} we validate the presented scheme 
by comparing thermodynamic properties obtained for the archetypical case of bcc Ti with available experimental data.
The method is, however, generally applicable, since an arbitrary energy surface can always be approximated by a piecewise quadratic function.
The computational tools and techniques used in the application of the method are described in Secs. \ref{sec:p4} and \ref{sec:result}.
\section{Piecewise Polynomial Potential Partitioning}{\label{sec:p4}}
\subsection {Partitioning the phase space}
\begin{figure}[h!]
\begin{center}
\includegraphics[width=0.95\columnwidth]{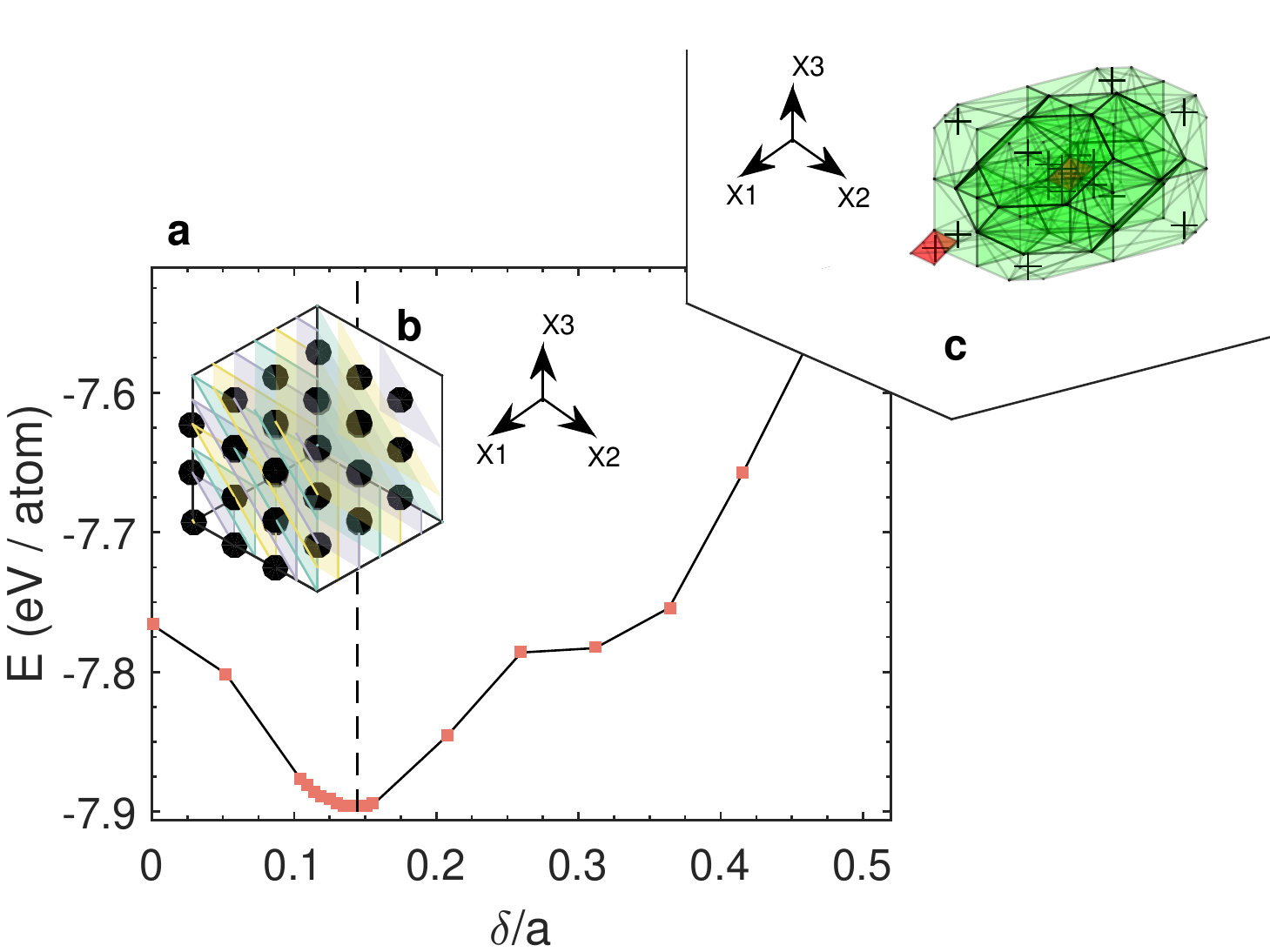} 
\caption{Locating the neighboring local minimum and creating augmented lattice. (a) Energy per atom versus distortion amplitude ($\delta$) along $\textbf{L}\frac{2}{3} (1,1,1)$ phonon of bcc Ti with lattice constant $a$. The volume of the supercell is the only degree of freedom which is equilibrated for each distortion. The minimum is located close to $\frac{\delta}{a}=0.1443$. (b) The (111) planes in a $\text{3}\times\text{3}\times\text{3}$ bcc supercell of bcc conventional unit cell, including 54 atoms. The three neighboring (111) planes are distinguished by different colors. (c) Representation of the augmented lattice unit cell and Voronoi cells containing lattice sites as generators. The + signs indicate lattice site in 3-dimensional space in periodic boundary conditions. The red cells are associated with bcc sites and the green cells are associated with corner sites.}
\label{fig:bcc2w}
\end{center}
\end{figure}

\begin{figure}[h!]
\begin{center}
\includegraphics[width=\columnwidth]{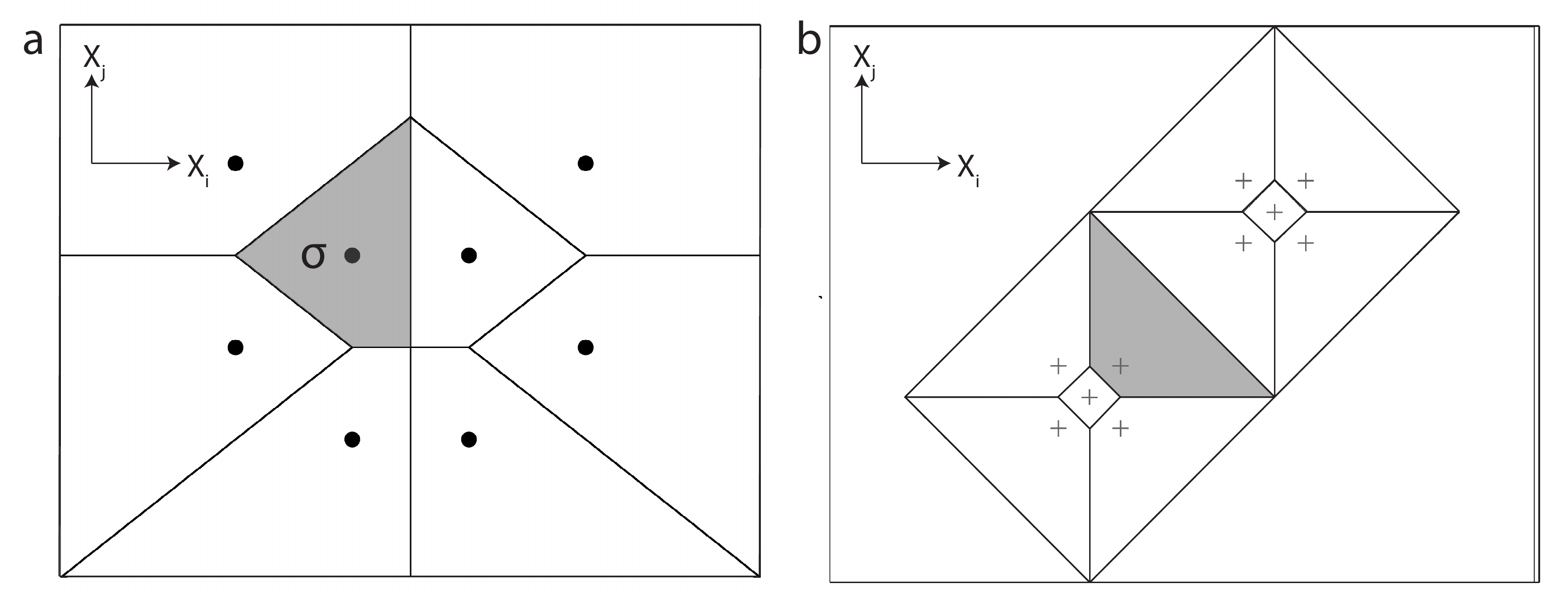} 
\caption{ Demonstration of partitioning the 3$N$-dimensional vs two-dimensional configuration space. (a) Schematic representation of partitioning 3$N$-dimensional  configuration space into its Voronoi tessellations. The generating points (circles) are 3$N$-dimensional position vectors. The shaded area includes all the coordinates that are associated with configuration $\sigma$. (b) Schematic 2D representation of partitioning the configuration space into its Voronoi tessellations. The + signs indicate the lattice sites that are the generating points in the Voronoi tessellation. The configurational state of simultaneous lying of two or more atoms in the shaded area does not belong to any $\zeta_{\sigma}$. Only those states with one atom in each Voronoi cell have a corresponding $\zeta_{\sigma}$.}
\label{fig:3d}
\end{center}
\end{figure}

To account for hopping of the system between different local distortions of the bcc lattice (see Fig. \ref{fig:multiple_local_min}), we construct an \textit{augmented lattice}, denoted by $L_{aug}$, that includes not only the ideal high-symmetry sites, but also the sites corresponding to configurations that are local energy minima near the ideal structure. These additional sites can be found by the identification of unstable modes through a standard lattice dynamics analysis, followed by a full relaxation calculation using, as an initial configuration, bcc structures slightly distorted along the various unstable modes.
Once one such local minimum has been identified, all other symmetrically equivalent minima can be identified by applying the symmetry operations of the high-symmetry phase (here, $O_h$ for bcc).
\\
In the case of bcc Ti, the unstable modes consist of longitudinal $[\xi \xi \xi]$ phonons with $\xi=2/3$~\cite{instab-dyn-1}. Geometrically, this mode moves two of the three neighboring (111) planes towards each other, whereas every third plane stays at rest. The collapse of the two planes results in the $\omega$ phase (for a displacement of $a\sqrt{3}/2$); therefore, this mode is associated with bcc to $\omega$ phase transition.
As shown in Fig. \ref{fig:bcc2w}(a), the potential energy for different distortion amplitudes along $\textbf{L}\frac{2}{3} (1,1,1)$ phonon is calculated by using first-principles electronic structure calculations.
All electronic structure calculations are performed by using the Vienna \textit{ab initio} simulation package ({\sc{VASP}})~\cite{vasp1,vasp2,vasp3,vasp4}, implementing the projector-augmented wave (PAW) method~\cite{paw}. The Perdew–Burke–Ernzerhof (PBE) functional~\cite{pbe} is used, with a plane-wave kinetic-energy cutoff of 222.3 eV 
for bcc Ti.
A $(\text{4}\times\text{4}\times\text{4})$ Monkhorst Pack mesh~\cite{monkhorst-pack} is used for generating the \textit{k}-space grid. For each distortion, the only degree of freedom equilibrated is the volume of the cell and the cell shape is fixed, because those structures that are combined to build the augmented lattice must be compatible. 
Volume relaxation for each supercell is performed in two steps (relaxing the initial structure and re-relaxing), using Methfessel Paxton method~\cite{methfesel}, with $\text{ISMEAR}=1$ and $\text{SIGMA}=0.15$. To obtain the energy (i.e., the energy of the relaxed volume structure), the tetrahedron method~\cite{tetrahedron} with $\text{ISMEAR}=-5$ is used.
\\ 
The minimum along the energy-distortion curve [as shown in Fig. \ref{fig:bcc2w}(a)], which corresponds to the $\omega$ structure, is the location of the first off-bcc site. This particular position of atoms is denoted by $\textbf{x}_{\omega}$, and the ideal bcc position is denoted by $\textbf{x}_{bcc}$. The augmented lattice supercell is constructed by operating the symmetry point group of bcc ($O_h$ point group) to a lattice including both sites at $\textbf{x}_{\omega}$ and $\textbf{x}_{bcc}$. As shown in Fig. \ref{fig:bcc2w}(c), this results in an augmented lattice with bcc lattice vectors and a basis consisting of nine sites;
the ideal bcc site, as well as eight other sites, each of which is located on a corner of a cube centered on a bcc site.
For the movement of atoms to represent the $\textbf{L}\frac{2}{3} (1,1,1)$ phonon, which involves the movement of three sets of $(111)$ planes, the supercell should, at least, consists of three sets of (111) planes or a third multiple of (111) planes. This is necessary because of the periodic boundary condition. In our calculations, a $3\times3\times3$ supercell of the bcc conventional unit cell is used, as presented in Fig. \ref{fig:bcc2w}(b). The lattice constant used for the conventional unit cell is $a=3.2413\text{\AA}$. This results in 54 bcc sites, therefore our simulation cell consists of $54\times9=324$ sites, including both bcc and corner sites.
\\
By creating an augmented lattice, phase space is partitioned into different \textit{configurations}. A \textit{configuration} $\sigma$ is defined as a possible assignment of Ti atoms and vacancies (Vac) to the augmented lattice sites. 
The configurational part of free energy associated with $L_{aug}$ is given by a 3$N$-dimensional integral over the classical configurational partition function in a ``coarse-grained'' form~\cite{vandewalle2002}:
\begin{equation}\label {config-partion-mp}
F_L=-k_BT\ln\sum_{\sigma \in L_{aug}}e^{-\beta F^*_{\sigma}}
\end{equation}
where 
\begin{equation}\label {define-constriant-vib-f}
F^*_{\sigma}=-k_BT\ln\int_{\textbf{x}\in\zeta_{\sigma}} e^{-\beta V(\textbf{x})}d\textbf{x} 
\end{equation}
where $\beta=1/(k_BT)$, $T$ is temperature, $k_B$ is Boltzmann's constant,  $\textbf{x}$ is a 3$N$ vector of all atomic positions, $N$ is the number of atoms in the system, $\zeta_{\sigma}$ is the proximity of configuration $\sigma$, which will be defined more precisely below, and $V(\textbf{x})$ is the potential energy of the system at a state represented by the position vector $\textbf{x}$.
\\
The above integration can be divided into two levels. One level is the ``outer'' level that sums over different configurations associated with $L_{aug}$ [represented in Eq. \eqref{config-partion-mp}], and, the other is the ``inner'' level that is a continuous integration of the Boltzmann distribution in the vicinity of each configuration [represented in Eq. \eqref{define-constriant-vib-f}]. The inner- level states are characterized by \textit{constrained vibrational free energy}, denoted by $F^*_{\sigma}$.  
The outer-level states can be sampled through a cluster-expansion approach~\cite{Sanchez84}, while a continuous sampling in the vicinity of each configuration is carried out by using the harmonic approximation of the energy hypersurface about each configuration.
\\
One has to define the proximity of each configuration for continuous sampling of the phase space. For each configuration $\sigma$, the unrelaxed position of the atoms is denoted by $\textbf{x}^u_{\sigma}$ (for each $\sigma$, $\textbf{x}^u_{\sigma}$ is a $3N$ vector). All the atomic positions closer to $\textbf{x}_{\sigma}^u$ than the unrelaxed position of any other configuration $\sigma'$, are defined as the proximity of configuration $\sigma$ and denoted by $\zeta_{\sigma}$. For computational efficiency, we determine $\zeta_{\sigma}$ by computing the Voronoi tessellation~\cite{voronoi} in tridimensional space generated by the augmented lattice points (which coincides with the well-known concept of Wigner Seitz cells~\cite{seitz}). These Voronoi cells are represented in Fig. \ref{fig:bcc2w}(c) for different augmented lattice points as generators. We then define $\zeta_{\sigma}$ in the 3$N$-dimensional configuration space as the Cartesian product of the tridimensional Voronoi cell associated with each site.
\\
This construction is preferable to a Voronoi tessellation in 3$N$-dimensional space, not only for computational reasons, but also because it naturally excludes nonphysical very-high-energy states from the region $\zeta_{\sigma}$, such as those where two atoms would lie in the same Wigner Seitz cell of a given lattice site. 
The contribution of the excluded states to the partition function is negligible, since these states are those of very high energy. Fig. \ref{fig:3d}(a) illustrates partitioning of 3$N$-dimensional configuration space schematically [to be compared with Voronoi tessellation in tridimensional space used to define $\zeta_{\sigma}$ as shown in Fig. \ref{fig:3d}(b)], and Fig. \ref{fig:3d}(b) indicates a schematic state which is implicitly excluded in our calculations due to the way $\zeta_{\sigma}$ is defined.
\subsection{Piecewise polynomial potential}
A piecewise polynomial form is employed in order to model the potential energy $V(\mathbf{x})$ over different subregions, $\zeta_{\sigma}$. We denote the location of the minimum of $V(\textbf{x})$ within $\zeta_{\sigma}$ by $\textbf{x}^r_{\sigma}$ (for each $\sigma$, $\textbf{x}^r_{\sigma}$ is a $3N$ vector). A harmonic expansion about $\textbf{x}^r_{\sigma}$ is used to calculate $F^*_{\sigma}$ in this case, although any order of polynomial is generally applicable. To determine the minimum within $\zeta_{\sigma}$ (which is a constrained nonlinear optimization procedure) a \textit{steepest-descent} search is performed and an interior or a boundary minimum is found.  
$\textbf{x}^u_{\sigma}$ is used as an initial guess, and a sequence of position vectors ($\textbf{x}^u_{\sigma}, \textbf{x}_1,\textbf{x}_2,. . .,\textbf{x}_{n+1} $) is constructed by moving along the negative gradient of potential-energy hypersurface by using the following equation:
\begin{equation}\label{sd}
\textbf{x}_{n+1}=\textbf{x}_{n}-\gamma_n \nabla V(\textbf{x}_n)
\end{equation}
where $\gamma_n$ is an arbitrary small scalar. 
\\
In the steepest-descent method, potential energy $V(\textbf{x}_n)$ and the gradient of potential energy $ \nabla V(\textbf{x}_n)$, are calculated by using {\sc{VASP}}. For each step, a value of 0.02 $\text{\AA}^2/\text{eV}$ is used for $\gamma_n$. If the movement along the negative gradient directs the configuration outside $\zeta_{\sigma}$, then the projection of the force vector, $\textbf{f}_v(\textbf{x}_n)$, along the boundary of  $\zeta_{\sigma}$ replaces the negative $\nabla V(\textbf{x}_n)$ in Eq. \eqref{sd}.
\\
If there exists no interior local minimum within $\zeta_{\sigma}$ (a positive-definite Hessian is the necessary but not the sufficient condition for the minimum to be an interior minimum), then  the minimum must be on the boundary of $\zeta_{\sigma}$. 
The first derivative may not vanish at a boundary minimum. 
The energy surface is expanded up to the quadratic term, with first and second derivatives of energy calculated at $\textbf{x}^r_{\sigma}$ (see supplementary note 1~\cite{sup}). 
All of the energies, forces and second derivatives of energy (force constant tensors) used in the piecewise polynomial model are calculated at $\textbf{x}^r_{\sigma}$ for each configuration employing PAW-PBE in {\sc{VASP}}. The supercell size ($\text{3}\times\text{3}\times\text{3}$ supercell of bcc unit cell) and $k$ points ($\text{2}\times\text{2}\times\text{2}$ Monkhorst Pack mesh) are chosen in a way that the energy values are converged with an accuracy of 1 meV/atom. The forces and second derivatives are converged with an accuracy of 1 meV${\text{\AA}}^{-1}$$\text{atom}^{-1}$ and 1 meV$\text{\AA}^{-2}$$\text{atom}^{-1}$, respectively.
\\
\subsection{Adiabatic switching technique}
As indicated in Eq. \eqref{define-constriant-vib-f}, the 3$N$-dimensional integral is over a complex hyper-volume $\zeta_{\sigma}$. Therefore, the analytic integration of the configurational partition function is not feasible, although the integral has known analytical solutions over certain simpler domains.
Moreover, divergence issues in the calculation of free energy due to existence of unstable modes present computational obstacles. Theses obstacles, however, can be bypassed by first constraining the integration domain to $\zeta_{\sigma}$ and then using the well-known adiabatic switching technique to calculate the associated constrained free energy. 
\\
The idea is to select a reference potential $V^{ref}(\textbf{x})$ with a positive-definite Hessian that is sufficiently stiff so that the integral in Eq. \eqref{define-constriant-vib-f} has the same value whether or not the integral is constrained over the domain $\zeta_{\sigma}$. The free energy of such a system, which is defined as $F^*_0$ in the following equation, can then be calculated analytically.
To obtain the free energy difference between the real and the reference systems, we need to define a thermodynamic path between these two states:
$\hat{V}(\textbf{x}, \lambda)=\hat{V} (\textbf{x},0)+ \lambda(\hat{V} [\textbf{x},1)-\hat{V} (\textbf{x},0)]$, which smoothly interpolates between the real [$\hat{V} (\textbf{x},1)=V(\textbf{x})$] and the reference [$\hat{V} (\textbf{x},0)=V^{ref}(\textbf{x})$] energy surfaces.
The free energy difference is derived to be the integration of ensemble-averaged potential-energy difference between the real and reference energy surfaces over $\zeta_{\sigma}$ along this path (see supplementary note 2 for derivation~\cite{sup}).
\begin{equation} \label{adiabticInt}
F^*_{1}-F^*_{0}=\int_0^1\left<\hat{V} (\textbf{x},1)-\hat{V} (\textbf{x},0)\right>_{\lambda}d\lambda
\end{equation}
where $\left<\hat{V} (\textbf{x},1)-\hat{V} (\textbf{x},0)\right>_{\lambda}$ denotes an average obtained through Metropolis sampling of the region $\zeta_{\sigma}$ using the potential-energy $\hat{V}(\textbf{x}, \lambda)$. $F^*_{1}$ for each subregion $\zeta_{\sigma}$ corresponds to the constrained vibrational free energy for the real system.
\subsection{Lattice gas model}
\begin{figure}[h!]
\begin{center}
\includegraphics[width=\columnwidth]{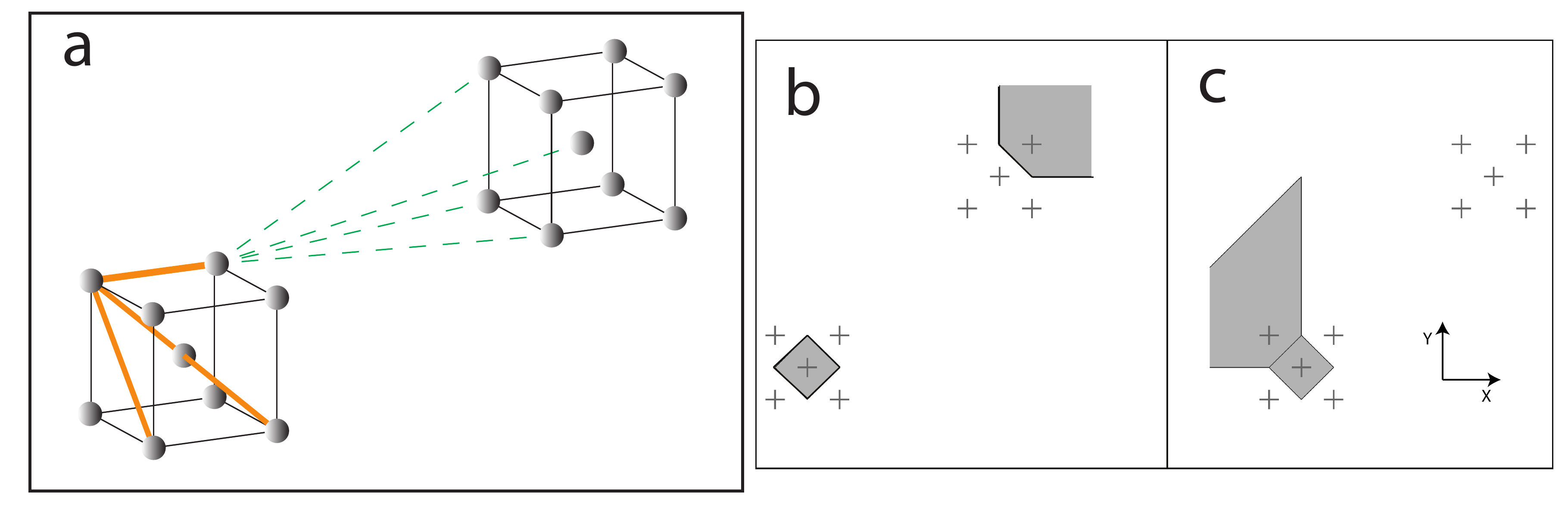} 
\caption{Demonstration of pairs included in the cluster expansion of vibrational free energy. (a) The bold red lines indicate short-pairs and the dashed green lines represent the long pairs. Not all of the long pairs considered in the cluster expansion are indicated, but only a few of them are represented as an example. A \textit{group} of sites is the collection of nine spheres sitting on the center and corners of a cube. Short pairs include the half-diagonal, edge, face-diagonal and diagonal pairs formed in the cube.  By imposing large ECIs for short pairs, they are avoided during a Monte Carlo simulation, eliminating the need to perform \textit{ab initio} energy calculations of those sates where more than one atom occupies a group of sites. (b) A two-dimensional plot of configuration space associated with a state during a Monte Carlo simulation. The shaded area is all the possible coordinates that the state can possess so that it is associated with a configuration in which Ti atoms occupy those sites in the center of shaded areas. (c) A two-dimensional representation of an example of a state that is precluded in Monte Carlo simulations due to large imposed ECIs for short pairs. The shaded areas indicate the coordinates corresponding to the state, where Ti atoms sit on sites on their centers.}
\label{fig:clusters}
\end{center}
\end{figure}

With the aid of commonly-used cluster expansion technique, constrained vibrational free energy corresponding to each configuration is represented as a polynomial series
in terms of the occupation variables $\sigma_i$ associated with each atomic site $i$ of the augmented lattice ($\sigma_i=+1$ if site $i$ is occupied by a Ti atom and $\sigma_i=-1$ if site $i$ is empty).
\begin{equation}\label{cl}
\frac{F^*(T,\sigma)}{N_s}=\sum_{\alpha} m_{\alpha} J_{\alpha}(T) \left< \prod_{i \in \alpha}\sigma_{i}\right>_{\alpha'}
\end{equation}
The sum in Eq. \eqref{cl} is over symmetrically distinct clusters $\alpha$ while the average is over clusters $\alpha'$ that are symmetrically equivalent to $\alpha$. $N_s$ is the number of sites in the parent lattice, $m_{\alpha}$ is the multiplicity of cluster $\alpha$, and $J_{\alpha}(T)$ is the effective cluster interaction (ECI) of cluster $\alpha$ at temperature $T$ (to be determined by a fitting procedure)~\cite{atat1}. Here, the effective cluster interactions are temperature-dependent, as opposed to the well-known cluster expansion of energy with temperature-independent ECIs~\cite{vandewalle2002}.\\
A cluster expansion has to be fit over a training data set. In this case, data points are the constrained vibrational free energy for a number of configurations, which are computed using the $\text{P}^\text{4}$ scheme. 
The expansion in Eq. \eqref{cl} is improved by adding more data points to the relatively small initial training data set until convergence is obtained in free energy value with a precision of a few meVs (see supplementary note 3 for a detailed description~\cite{sup}).
In our calculations, 34 different clusters are included. There is one empty cluster (which is the constant term in the polynomial series) and there are two point clusters (one is a bcc point cluster and one is a corner point cluster), along with 31 pair clusters. As shown in Fig. \ref{fig:clusters}(a), four of these pair clusters connect the sites within each group of nine sites (one bcc site and eight neighboring corner sites) and are accordingly called ``short-pairs''. The other 27 pairs join one site on a group to another site in the nearest-neighbor group .
The four short-pairs include a half diagonal pair, an edge pair, a face diagonal pair, and a diagonal pair of the cube formed by eight corner sites and one bcc site in the center. 
All of the ECIs associated with these clusters are fit by using the training \textit{ab initio} data set, except for those ECIs associated with short-pairs. These four pairs are treated differently in our scheme because their role is simply to ensure that two atoms never lie within the same group of nine sites. Such configurations have an energy so high that they essentially never occur in reality. To avoid performing \textit{ab initio} calculations for these unphysical states, these ECIs are simply set to a positive value sufficiently large to effectively suppress the appearance of more than one atom within the same nine-site group of sites. A two-dimensional plot of such an unphysical state is represented in Fig. \ref{fig:clusters}(c) and is compared with those states included in our calculation; a typical example of them is presented in Fig. \ref{fig:clusters}(b).
A set of temperature-dependent ECIs, $\{J(T)\}$, is fit over a training data set at different temperatures employing the alloy-theoretic automated toolkit (ATAT)~\cite{atat1,atat2,atat3} [by Eq. \eqref{cl}].
\\
The $\text{P}^\text{4}$ scheme is employed only once, at a temperature $T_1$, for each configuration to calculate $F^*(\sigma,T_1)$.
Constrained vibrational free energy at any other temperature $T_2$ is simply calculated by knowing $F^*$ at $T_1$ by using the well-known thermodynamic integration, reformulated as follows:
\begin{equation}\label{extend_T} 
\frac{F^*(T_2,\sigma)}{T_2}=\frac{F^*(T_1,\sigma)}{T_1}+\int_{T_1}^{T_2}\left<U_{\sigma}(T)\right>d(1/T),
\end{equation}
where $\left<U_{\sigma}(T)\right>$ is the average internal energy of the system constrained to $\zeta_{\sigma}$, which is calculated using Metropolis sampling (see supplementary note 5 for the derivation of Eq. \eqref{extend_T}~\cite{sup}). 
In other words, once a data set of $F^*$ is built at a temperature $T_1$, the corresponding data set at any other temperature $T_2$ is calculated by using $F^*(T_1,\sigma)$ as an initial point of integration in Eq. \eqref{extend_T}.
\\
The thermodynamic integration technique makes the calculation of total free energy in Eq. \eqref{config-partion-mp} feasible. To calculate the Helmholtz free energy at any temperature $T_1$, we employ thermodynamic integration in accordance with the following equation:
\begin{equation}\label{canonical_int}
\frac{F(T_1)}{T_1}-\frac{F(T_0)}{T_0}=\int_{T_0}^{T_1}\left<F^*(T)\right>d(1/T),	
\end{equation}
where $\left<F^*(T)\right>$ is the average vibrational free energy at temperature $T$ and $F(T)$ is the total free energy at $T$ (see supplementary note 5~\cite{sup}). 
A Monte Carlo simulation is carried out to compute the ensemble average of constrained vibrational free energy, $\left<F^*\right>$, with Eq. \eqref{cl} once the set of $\{J(T)\}$ is known, utilizing the multicomponent easy Monte Carlo code ({\sc{memc2}})~\cite{atat3}.
We need a convenient initial point of integration in Eq. \eqref{canonical_int}, whose free energy can be computed analytically. Therefore, the mean-field approximation limit at high temperature is used as the starting point for integration~\cite{vandewalle_self}. The Helmholtz free energy at high temperatures is calculated analytically (Supplementary note 4~\cite{sup} gives a detailed description of the mean-field approximation (MFA) calculation). Any thermodynamic process that connects the initial state to the final state can be integrated to obtain the free energy. 
To avoid calculating $\left<F^*(T)\right>$ at temperatures far from the temperature range of interest, the ECIs are kept constant (at their value at the desired temperature $T_1$) along the thermodynamic integration path while the temperature the system experiences varies from our high-temperature reference down to the temperature of interest [which is why the integrand in Eq. \eqref{canonical_int} is $\left<F^*(T)\right>$ rather than $\left<U^*(T)\right>$]. Only at the end of the integration path does the free energy represent the one of a real physical system. 
\\
To include the thermal electronic contribution to the free energy, the Fermi distribution ($\text{ISMEAR}=-1$) is employed with the corresponding smearing parameter at each temperature $T$ ($\text{SIGMA}= k_BT$) in all of the \textit{ab initio} calculations. For each configuration $\sigma$, the electronic contribution $F_{elec,\sigma}$ to the total free energy is added to the corresponding vibrational free energy $F^*_{\sigma}$ at different temperatures. As a result, the free energies used in the training data set of Eq. \eqref{cl} to obtain the corresponding ECIs are $F^*_{\sigma}+F_{elec,\sigma}$ at each temperature $T$.
\\
The fact that we model the system's potential-energy surface by a piecewise approximation rather than by a smooth surface has no bearing on the accuracy of the calculated free energies. It can be shown that the error in the free energy is bounded by the largest error in the energy, regardless of the smoothness of the approximation. A proof of the above argument is presented in supplementary note 8~\cite{sup}. 
\section{Results}
\begin{figure}[h!]
\begin{center}
\includegraphics[width=\columnwidth]{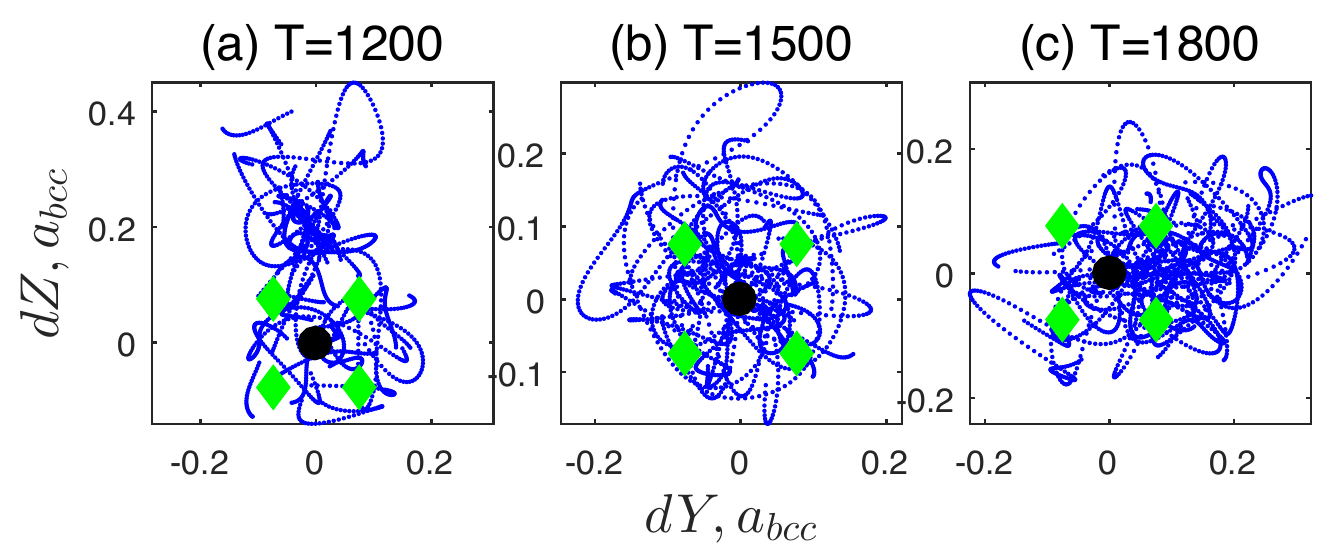}
\caption{Two-dimensional projection of the trajectory of a typical atom in the $\text{Ti}_\text{54}$ supercell during a MD simulation for (a) 1200 K, (b) 1500 K, and (c) 1800 K. Displacements are normalized by the lattice constant, $a_bcc$. Only the positive half of the $x$ axis is projected for the purpose of clarity. The black circle indicates the ideal bcc position and the green diamonds are the off-bcc sites augmented to the ordinary bcc lattice in the $\text{P}^\text{4}$ scheme.}
\label{fig:hop1}
\end{center}
\end{figure}

\begin{figure}[h!]
\begin{center}
\includegraphics[width=\columnwidth]{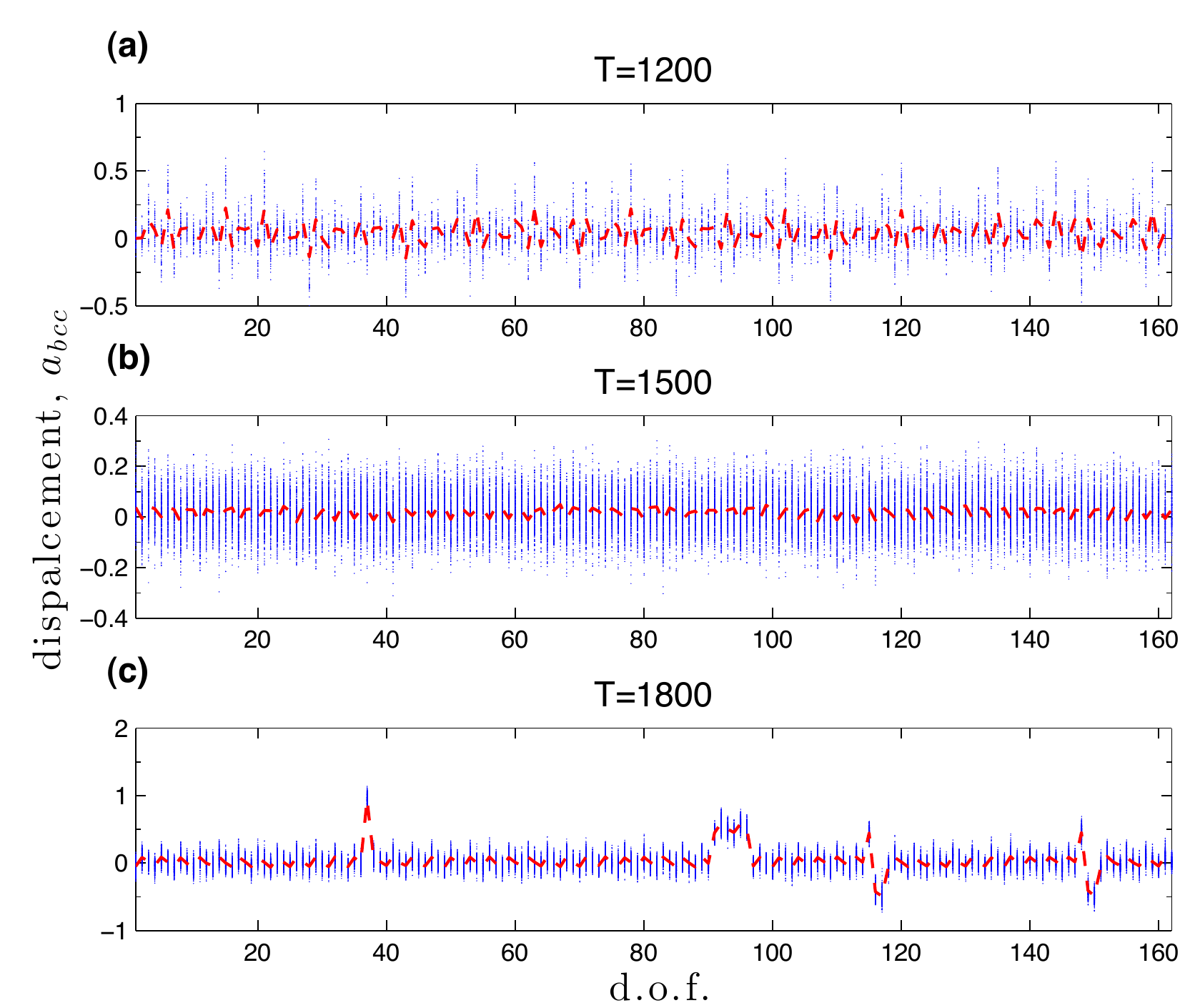}
\caption{Displacement from ideal bcc position normalized by lattice constant for each degree of freedom (d.o.f.) obtained from molecular dynamics (MD) trajectories in a fixed bcc $\text{Ti}_\text{54}$ supercell at (a) 1200 K, (b) 1500 K, and (c) 1800 K. The average displacements for each d.o.f. are connected by a red line.}
\label{fig:hop2}
\end{center}
\end{figure}

\begin{figure}[h!]
\begin{center}
\includegraphics[width=\columnwidth]{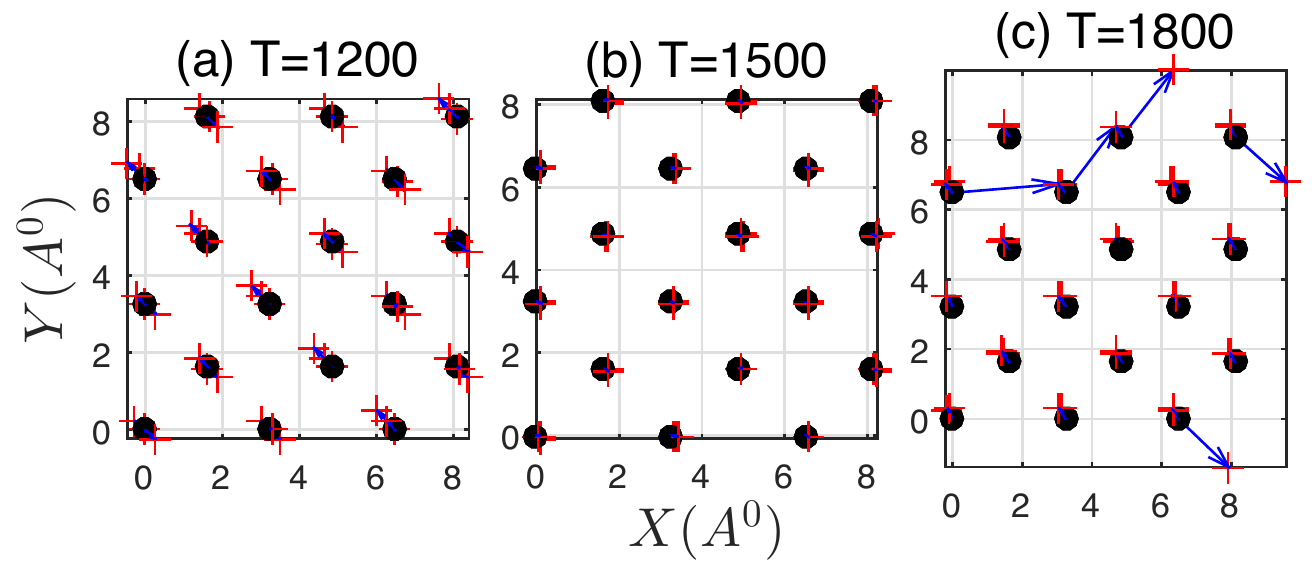}
\caption{Two-dimensional plot of MD time-averaged atomic positions (red plus signs) and the ideal bcc positions (black circles) for (a) 1200 K, (b) 1500 K, and (c) 1800 K. The deviation from bcc positions are indicated by arrows. All atomic positions are in $\text{\AA}$ units.}
\label{fig:hop3}
\end{center}
\end{figure}

\begin{figure}[h!]
\begin{center}
\includegraphics[width=\columnwidth]{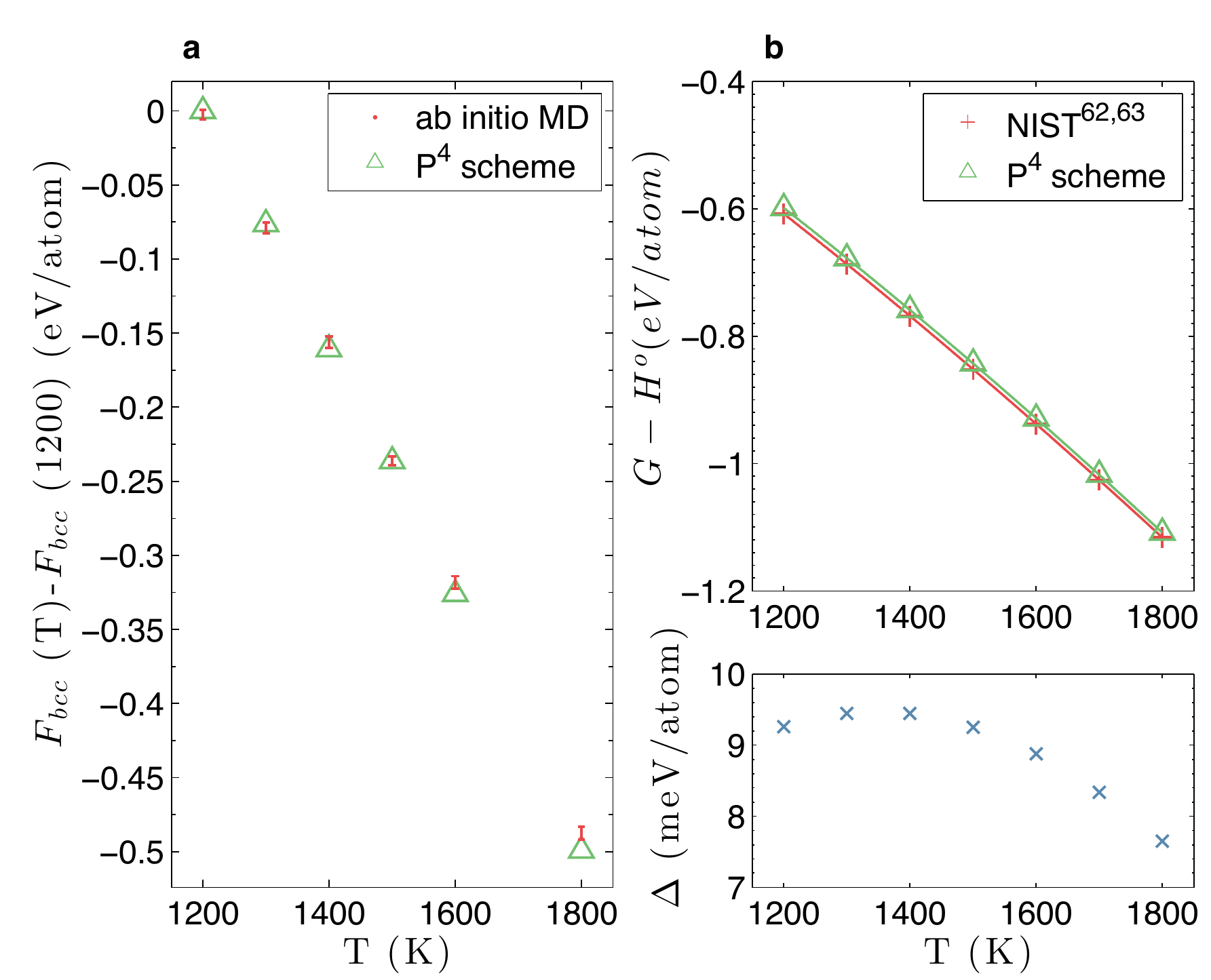} 
\caption{(a) Helmholtz free energy of bcc Ti with respect to its free energy at 1200 K compared with molecular dynamics (MD) results for internal consistency check. A $\text{Ti}_\text{54}$ supercell is used both in MD and $\text{P}^\text{4}$ at all temperatures. Electronic contribution to the free energy is included. Error bars indicate the accuracy of MD values for free energy. The accuracy of $\text{P}^\text{4}$ results are ensured to be less than 1 meV/atom. (b) Gibbs free energy of bcc Ti with respect to enthalpy of hcp Ti at room temperature. The curve indicates a function of the form $G-H^0=a+bT+cT\ln(T)+dT^2$ which is fit to the $\text{P}^\text{4}$ values. The fit coefficients are $a=0.0160155$, $b=9.586\times10^{-4}$, $c=1.984\times10^{-4}$ and $d=-5.331\times10^{-8}$. $\Delta$ is the difference between the calculated value and NIST data and is less than 10 meV/atom at all temperatures.}
\label{fig:NISTMD-2}
\end{center}
\end{figure}

\begin{figure}[h!]
\begin{center}
\includegraphics[width=\columnwidth]{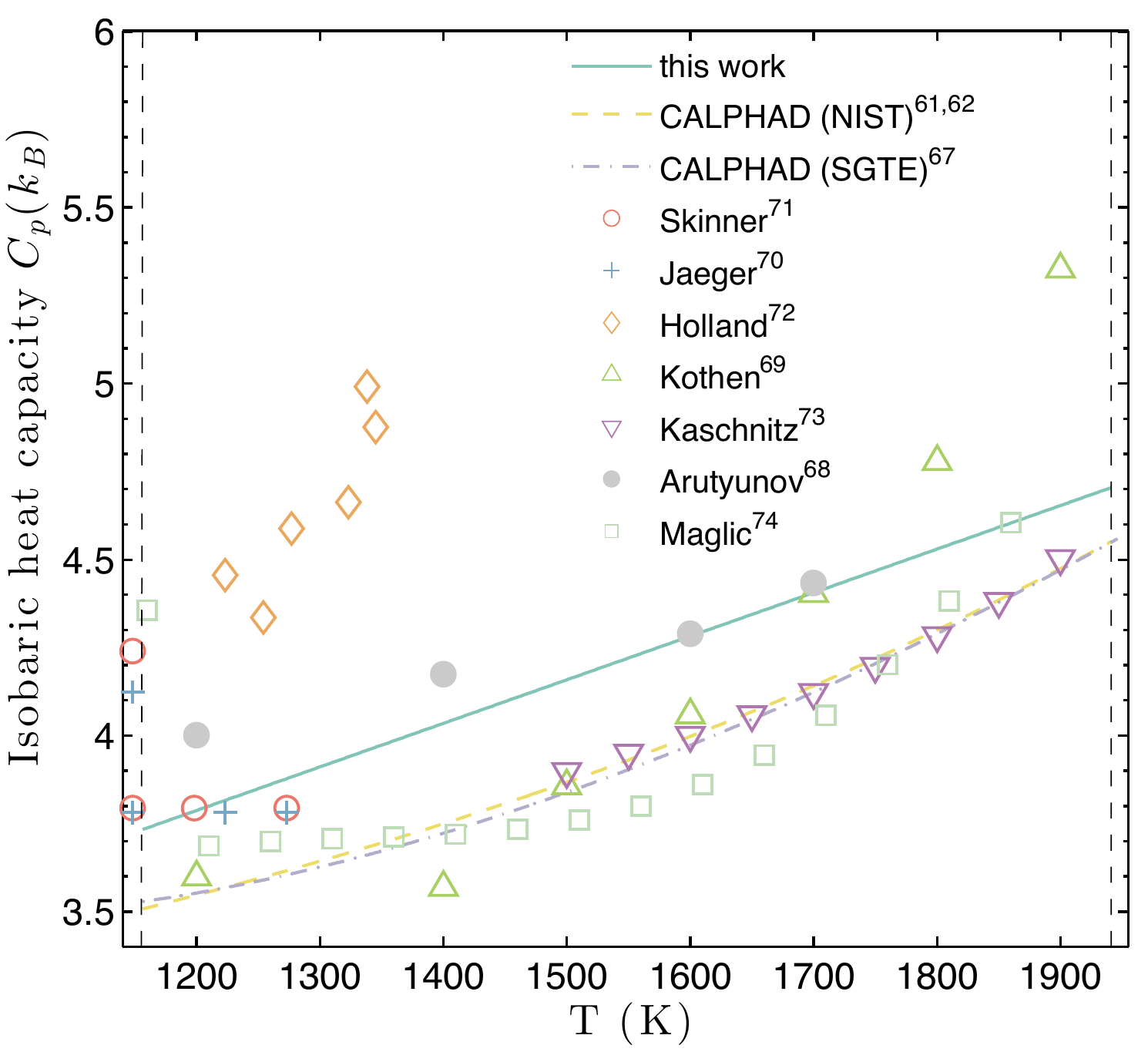} 
\caption{Isobaric heat capacity of bcc Ti at ambient pressure (0.001 kbar). The presented s scheme is used to calculate the heat capacity of  the mechanically unstable phase of bcc. The vertical lines indicate the experimental hcp-to-bcc transition temperature, $T^{\text{hcp}\rightarrow\text{bcc}}_{\text{exp}}=1156.15$ K (Ref. ~\cite{instab-dyn-1}), and melting temperature, $T^{\text{melt}}_{\text{exp}}=1933.15$ K (Ref. ~\cite{instab-dyn-1}), respectively. {\sc{CALPHAD}} data are taken from both SGTE (Ref. ~\cite{SGTE}) and NIST (Ref. ~\cite{nist-1,nist-2}). Experimental values are presented from Ref. ~\cite{Arutyunov,Kothen,Jeager,skinner,Holland,Kaschnitz,Maglic}.}
\label{fig:cp}
\end{center}
\end{figure}

\begin{figure}[h!]
\begin{center}
\includegraphics[width=\columnwidth]{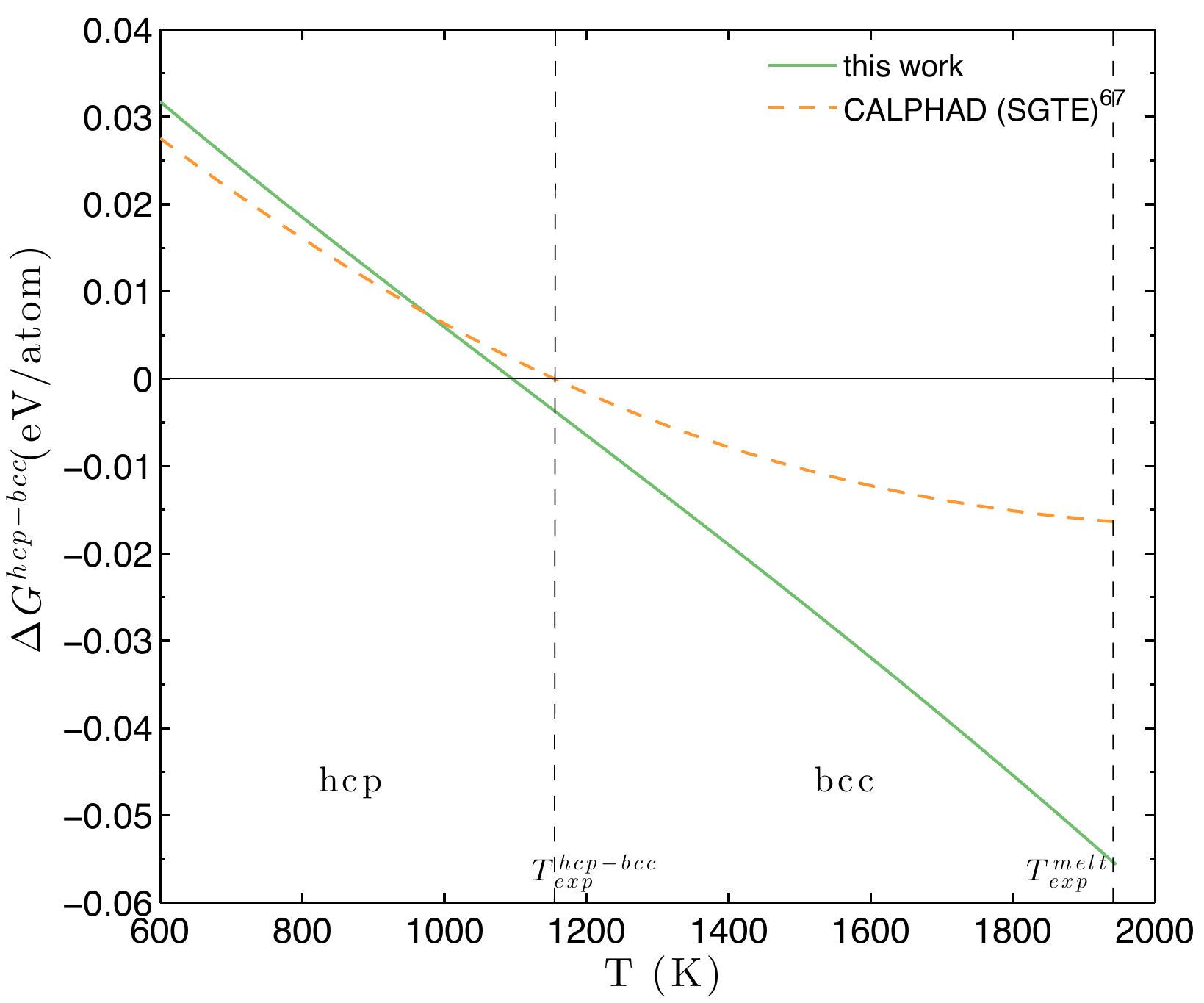} 
\caption{Gibbs free energy difference between bcc and hcp phase of Ti, $\Delta G^\text{hcp-bcc}$ at ambient pressure (0.001 kbar).The vertical lines indicate the experimental hcp-to-bcc transition temperature $T^{\text{hcp}\rightarrow\text{bcc}}_{\text{exp}}=1156.15$ K (Ref.~\cite{instab-dyn-1}) and melting temperature $T^{\text{melt}}_{\text{exp}}=1933.15$ K (Ref. ~\cite{instab-dyn-1}). {\sc{CALPHAD}} values are taken from the SGTE unary database (Ref. ~\cite{SGTE}).}
\label{fig:TT}
\end{center}
\end{figure}

{\label{sec:result}}
The free energy of bcc Ti 
is calculated by using the $\text{P}^\text{4}$ method described in the previous section.
To validate the obtained free energy, we compare our results with \textit{ab initio} molecular dynamics (MD) and National Institute of Standards and Technology (NIST) condensed phase experimental thermodynamic data for Ti 
~\cite{nist-1,nist-2}. 
To first perform an internal consistency check of the method, a set of NVT molecular dynamics calculations are performed at different temperatures for a $\text{3}\times\text{3}\times\text{3}$ supercell of bcc conventional unit cell, including 54 atoms, using {\sc{VASP}}~\cite{vasp1,vasp2,vasp3,vasp4}. 
In constant temperature \textit{ab initio} MD simulations, the thermostat is conducted under Nos$\acute{\text{e}}$-Hoover chain formalism~\cite{nose-1,nose-2,nose-3,nose-4}, and ensemble averages are captured every 50 steps (150 fs) in a trajectory with sufficient number of steps which ensures an accuracy of 5 meV/atom. The internal energy calculated in the NVT \textit{ab initio} MD simulations includes both the kinetic and potential energies. The electronic contribution to the total energy is considered by using Fermi smearing ($\text{ISMEAR}=-1$) along with the corresponding smearing parameter at each temperature $T$ ($\text{SIGMA}=k_BT$).
\\
Before comparing the free energies, we use the MD results to illustrate the hopping of the system around local distortions of the bcc structure to verify the hypothesis used in the $\text{P}^\text{4}$ scheme.
The trajectory of a typical atom in the $\text{Ti}_\text{54}$ bcc supercell is illustrated in Figs. \ref{fig:hop1}(a)-\ref{fig:hop1}(c) for 1200, 1500, and 1800 K, respectively. It is observed that the trajectory at 1500 K is more symmetric compared with 1200 and 1800 K. The reason for this behavior is that, at low temperatures, the system preferably samples configurations that are similar to hcp. At very high temperatures, the atoms diffuse across different bcc sites with a high probability instead of remaining in the vicinity of one bcc site.
The displacement relative to the ideal bcc position for each degree of freedom (d.o.f) is indicated in Figs. \ref{fig:hop2}(a)-\ref{fig:hop2}(c) for 1200, 1500, and 1800 K, respectively. As shown in Fig. \ref{fig:hop2}, the average displacement for each d.o.f. is almost zero at 1500 K, which explains the symmetry observed in the atomic trajectory at this temperature. Although the average displacement is around zero, the system experiences displacement amplitudes in the order of 1/3 of nearest-neighbor (NN) pair distance at 1500 K. However, for 1200 and 1800 K, some degrees of freedom indicate larger and asymmetric displacements, which results in an off-zero average displacement for those d.o.f. Some of these asymmetric displacement amplitudes equal the lattice constant, $\text{a}_\text{bcc}$, for 1800 K, which implies some atomic permutations at this temperature.
This is further clarified in Fig. \ref{fig:hop3}, where the time-averaged position of each atom in the $\text{Ti}_\text{54}$ supercell is shown. 
For the MD simulation at 1500 K, the time-averaged structure is exactly bcc, as shown in Fig. \ref{fig:hop3}(b). However, for the MD simulation at 1200 K, which is just above the transition temperature to hcp, the time-averaged structure is slightly shifted toward the path that transforms bcc to hcp [see Fig. \ref{fig:hop3}(a)]. This explains the asymmetry observed in the atomic trajectory at this temperature. The averaged atomic positions at 1800 K stay at bcc, although some permutation of atoms are observed [as shown in Fig. \ref{fig:hop3}(c)], which is the result of high thermal energy accessible to the system at temperatures close to the melting point (1941 K). 
\\
We now compare the free energies obtained through the $\text{P}^\text{4}$ method with those obtained by \textit{ab initio} MD. The MD simulations are carried out to trace the average internal energy of bcc Ti 
at different temperatures. Helmholtz free energy is obtained by integrating the thermodynamic relation $$\left[\frac{\partial (F/T)}{\partial (1/T)}\right]_{V,N}=U,$$
where U is the average internal energy. 
The Helmholtz free energy of bcc Ti relative to 
 bcc Ti free energy at 1200 K, obtained with both the $\text{P}^\text{4}$ method and \textit{ab initio} MD, is illustrated in Fig. \ref{fig:NISTMD-2}(a). The agreement between the $\text{P}^\text{4}$ method and \textit{ab initio} MD confirms the validity of the presented scheme.
\\
The enthalpy of hexagonal closed-pack (hcp) Ti
 at room temperature (RT) is assigned as the reference point for energy to make our results comparable to NIST data, since hcp is the reference phase for titanium. Formation of the hcp phase in titanium from around 1150 K down to 0 K at ambient pressure implies the mechanical stability of hcp Ti, which is further confirmed by standard phonon analysis. Therefore, the free energy of hcp Ti is calculated by using a standard harmonic approximation utilizing the {\sc{fitfc}} code~\cite{atat1,atat3}. The harmonic model can be regarded as a special case of the $\text{P}^\text{4}$ scheme. When the $\text{P}^\text{4}$ scheme is applied to a mechanically stable phase, with the phase at the local minimum of the energy surface, it reduces to the harmonic model. In this case, the augmented lattice is basically the same as the ordinary lattice of hcp phase. Consequently, the summation in Eq. \eqref{config-partion-mp} reduces to one term and Eq. \eqref{define-constriant-vib-f} becomes the standard free energy equation of a harmonic model.   
\\
For calculating the force constant tensor for hcp Ti, the {\sc{fitfc}} code includes up to the third nearest-neighbor forces in a 90-atom supercell with 0.1 $\text{\AA}$ displacements.
Once the free energy and entropy of hcp Ti are calculated at room temperature, the enthalpy of hcp Ti
is computed by using the thermodynamic relation H=F+TS+pV, which is used as the reference point in our free energy results. Moreover, the pV work term correction is added to the computed Helmholtz free energy of bcc Ti in order to get Gibbs free energy at ambient pressure (see supplementary note 6 for more details~\cite{sup}). Available NIST values are also Gibbs free energies at ambient pressure with reference to hcp Ti enthalpy at room temperature. 
\\
As shown in Fig. \ref{fig:NISTMD-2}(b),
the $\text{P}^\text{4}$ method predictions for free energy of bcc Ti
at different temperatures agree perfectly well with NIST data. The discrepancy between this work and experimental data is less than a few meVs, which is the accuracy needed to resolve energy differences that typically drive solid-state phase transitions. 
\\
The data obtained for Gibbs free energy with respect to the reference state, G, are used to fit a power series of the following form 
\begin{equation}\label{fitG} 
G\left(T\right)=a+bT+cT\ln\left(T\right)+dT^2
\end{equation}
where $a=0.0160155$, $b=9.586\times10^{-4}$, $c=1.984\times10^{-4}$ and $d=-5.331\times10^{-8}$ are coefficients that are fit through a least square fitting procedure. The constant-pressure heat-capacity curve is then obtained according to the relation $C_p\left(T\right)=-T\left(\frac{\partial^2 G(T,p)}{\partial T^2}\right)_p$ and is presented in Fig. \ref{fig:cp}. NIST and SGTE (Scientific Group Thermodata Europe) curves~\cite{SGTE}, along with the experimental values at different temperature ranges~\cite{Arutyunov,Kothen,Jeager,skinner,Holland,Kaschnitz,Maglic}, are compared with the $\text{P}^\text{4}$ isobaric heat-capacity curve. 
Our results for bcc Ti are seen to lie within the spread in the available experimental results. The observed discrepancy among different experimental measurements corresponds to the experimental accuracy. This is especially important for higher-order properties such as heat-capacity, which is usually measured with lower accuracy.
It is also important to note that the determination of derivative quantities such as heat-capacity would result in significant noise in such quantities, since they are highly sensitive to extremely small free energy differences. 
\\
As indicated in Fig. \ref{fig:TT}, by comparing the calculated Gibbs energies for hcp and bcc Ti, the transition temperature from hcp to bcc is obtained. The calculated transition temperature of 1095 K compares well with the experimental values of 1156.15 K~\cite{instab-dyn-1} and 1198.26 K~\cite{nist-1} and computational results of 1114 K~\cite{Debye} and 1250 K~\cite{MD}. 
As indicated in Fig. \ref{fig:TT}, our calculation shows a Gibbs energy difference between hcp and bcc that has a similar slope to {\sc{CALPHAD}} data at lower temperature (below and slightly above  the transition temperature). However, at higher temperature the slopes differ and our calculations show a steeper negative slope. This originates from using the harmonic approximation for the hcp Gibbs energy, which is insufficient to incorporate anharmonic contributions into the free energy. These contributions becomes crucial especially at high temperature. Ignoring anharmonic free energy for hcp Ti results in an overestimation of Gibbs energy of hcp and consequently a steeper slope at temperatures above the transition temperature. One should not, however, correlate all of the disagreement with the harmonic model deficiency considering that, at temperatures above transition to bcc, the {\sc{CALPHAD}} Gibbs energies of hcp are based on extrapolations of experimental data from a region of temperature where hcp is stable.
\section{Discussion}{\label{sec:discussion}}
It is important to observe that even though the method only requires \textit{ab initio} calculations of harmonic force constants, the resulting free energy model includes anharmonic contributions to an accuracy that can be systematically improved by simply including more sites in the augmented lattice.
Phonon lifetime effects, stabilization by quartic terms, etc. are all included in the thermodynamic description, despite the use of a local harmonic treatment. This follows from the fact that any smooth function can be approximated with any given accuracy by a piecewise quadratic function, provided the pieces are chosen sufficiently small. It is practically more convenient to improve an approximation to a function by including more polynomial pieces of a low order (e.g., quadratic) than to increase the order of a single polynomial, as effective Hamiltonian methods traditionally do. The former avoids Runge's phenomenon~\cite{Epperson} while the latter does not. Figure 2 in supplementary notes~\cite{sup} represents the piecewise polynomial interpolation of an illustrative potential-energy surface. In addition, in our method, when the expansion points of the quadratic pieces are chosen to lie at local minima, the approximation is most accurate in the regions of phase space where the system spends the most time, a property not guaranteed by a higher-order polynomial expansion of the system's energy surface.
\\
Our method becomes especially advantageous relative to existing methods, such as thermodynamic integration (TI) methods, when the anharmonicity is so strong that it creates multiple local minima around a local high-symmetry maximum. 
In such cases, thermodynamic integration can become problematic because the hops between the local minima can become rare, resulting in difficulties in equilibrating the system at each step of the thermodynamic integration. The residence time in a well grows exponentially with the well depth, so even with an efficient energy model, thermodynamic integration may be too computationally demanding.
In such cases, sampling through \textit{ab initio} molecular dynamics (AIMD) would become inefficient and impractical whereas Monte Carlo sampling by means of a lattice gas model (as in the  $\text{P}^\text{4}$ method) remains equally efficient. 

TI methods calculate anharmonic free energy employing the quasiharmonic model as a reference. Therefore, the efficiency of TI methods greatly depends on how close the quasi-harmonic approximation is to the real system, while the efficiency of our method depends much less on this fact, because the lambda integration in the presented model switches between local harmonic approximations inside subregions. For strongly anharmonic systems, where the quasiharmonic sampling is far away from the \textit{ab initio} sampling, TI methods become less efficient. 
\section{Conclusion}{\label{sec:conclusion}}
In summary, a general and robust scheme to determine the free energy of mechanically unstable but dynamically stabilized phases is presented. The reasonable agreement between the free energy obtained for the prototypal example of bcc Ti and the existing experimental and computational values confirms the validity of our method. The proposed method also offers a natural avenue to handle alloys since the cluster expansion method can easily allow for multiple species on each site of the augmented lattice introduced herein. 

\begin{acknowledgments}
\small
This work is supported by the US National Science Foundation by Grants NO. DMR-1154895 and No. DMR-1505657 and by Brown University through the use of the facilities of its Center for Computation and Visualization. This work uses the Extreme Science and Engineering Discovery Environment (XSEDE), which is supported by the National Science Foundation Grant No. ACI-1053575.
\end{acknowledgments}

%
%

\end{document}